
\input phyzzx.tex
\input tables.tex

\def\etmiss{E_T^{miss}}
\def\mb{m_b}
\def\cnone{\widetilde\chi_1^0}
\def\sina{\sin\alpha}
\def\cosa{\cos\alpha}
\def\hl{h^0}
\def\hh{H^0}
\def\ha{A^0}
\def\hm{H^-}
\def\hp{H^+}

\def\mha{m_{\ha}}

\def\hp{H^+}
\def\gam{\gamma}
\def\mstop{M_{\wtilde t}}

\def\gev{~{\rm GeV}}
\def\tev{~{\rm TeV}}

\def\fbi{~{\rm fb}^{-1}}


\def\prdj#1{{\it Phys. Rev.} {\bf D{#1}}}
\def\npbj#1{{\it Nucl. Phys.} {\bf B{#1}}}
\def\prlj#1{{\it Phys. Rev. Lett.} {\bf {#1}}}
\def\plbj#1{{\it Phys. Lett.} {\bf B{#1}}}

\def\ibid{{\it ibid.}}
\def\mt{m_t}

\def\tauptaum{\tau^+\tau^-}
\def\rta{\rightarrow}
\def\tanb{\tan\beta}

\def\cosb{\cos\beta}

\def\lplm{l^+l^-}

\def\nsd{N_{SD}}

\def\mz{m_Z}
\def\anti{\overline}

\def\ifmath#1{\relax\ifmmode #1\else $#1$\fi}
\def\half{\ifmath{{\textstyle{1 \over 2}}}}

\def\3quarter{{\textstyle{3 \over 4}}}

\def\ebtag{e_{b-tag}}
\def\emistag{e_{mis-tag}}
\def\h{h}
\def\mh{m_{\h}}

\input phyzzx
\Pubnum={$\caps UCD-94-7$\cr $\caps SMU-HEP-94/4$ \cr
$\caps UCSD/PTH-94-2$\cr}
\date{March, 1994\cr Revised September, 1994\cr}

\titlepage
\vskip 0.75in
\baselineskip 0pt 
\hsize=6.5in
\vsize=8.5in
\centerline{{\bf LHC Detection of Neutral MSSM Higgs Bosons}}
\centerline{{\bf via $\bf gg\rta b\anti b \h\rta b\anti b b\anti b$}}
\vskip .075in
\centerline{J. Dai$^a$, J.F. Gunion$^b$ and R. Vega$^c$}
\vskip .075in
\centerline{\it a) Dept. of Physics, University of California at San Diego,
La Jolla, CA 92093}
\centerline{\it b) Davis Institute for High Energy Physics,
Dept. of Physics, U.C. Davis, Davis, CA 95616}
\centerline{\it c) Dept. of Physics, Southern Methodist University,
Dallas, TX 75275}

\vskip .075in
\centerline{\bf Abstract}
\vskip .075in
\centerline{\Tenpoint\baselineskip=12pt
\vbox{\hsize=12.4cm
\noindent We demonstrate that simultaneous detection of two
of the neutral Higgs bosons of the Minimal Supersymmetric Model
will be possible at the LHC in the $b\anti b b\anti b$ final
state, provided $\tanb$ is large and $b$-tagging can be performed
with good efficiency and purity. This mode enhances the
already established guarantee that
there is no region of supersymmetric parameter space
for which {\it none} of the Higgs bosons of the model
can be seen at the LHC.
}}

\vskip .15in
\noindent{\bf 1. Introduction and Procedure}
\vskip .075in

The establishment of complementary techniques and channels that
guarantee the detection of the Higgs bosons of the Minimal Supersymmetric
Model (MSSM) at the LHC has been an ongoing process. In early
\REF\everybody{J.F. Gunion and L.H. Orr, \prdj{46} (1992) 2052.
Z. Kunszt and F. Zwirner, \npbj{385} (1992) 3.
H. Baer, M. Bisset, C. Kao and X. Tata,
\prdj{46} (1992) 1067.
V. Barger, K. Cheung, R.J.N. Phillips and A.L. Stange,
\prdj{46} (1992) 4914.}
\REF\pergunion{For a review see, J.F. Gunion,
{\it Perspectives on Higgs Physics}, ed. G. Kane,
World Scientific Publishing (1992), p. 179.}
work,\refmark{\everybody,\pergunion} it was demonstrated that
the $\gam\gam$ and $4\ell$ decay channels of inclusively produced
CP-even $\hl$ and $\hh$ Higgs bosons and the $t\rta \hp b$ decays
of a top quark combine to yield viable Higgs
signals at the LHC for large regions of MSSM parameter space.
However, in general there remain regions of parameter space such
that detection of an MSSM Higgs boson will not be possible via these modes
at the LHC (and perhaps also not at LEP-200).
For example, if $\mstop\sim 1\tev$ and $\mt\sim150\gev$ there is a window
of $\mha$--$\tanb$ parameter space, with $110\lsim\mha\lsim 170\gev$
and $\tanb\gsim 10$, such that no MSSM Higgs boson would be easily found.
This window is reduced but not entirely absent for the CDF value of $\mt\sim
175\gev$.

Various additional techniques/channels for MSSM Higgs discovery have since been
explored.  Channels involving Standard Model particle decay modes
include the $\tauptaum$ decay channels of the $\hh$ and $\ha$,
\Ref\kz{Z. Kunszt and F. Zwirner, Ref.~[\everybody].}\
$t\anti t \hh$ and $t\anti t \hl$ detection in the $t\anti t b\anti b$
\REF\dgvi{J. Dai, J.F. Gunion, R. Vega, \prlj{71} (1993) 2699.}
\REF\dgvii{J. Dai, J.F. Gunion, R. Vega, \plbj{315} (1993) 355.}
final state,\refmark{\dgvi,\dgvii} $W\hl$ and $W\hh$
detection in the $Wb\anti b$
\REF\stangeetalwhsm{A. Stange, W. Marciano and S. Willenbrock,
\prdj{49} (1994) 1354, and preprint ILL-TH-94-8 (1994).}
final state,\refmark\stangeetalwhsm\
and $t\anti b \hm,\anti t b\hp$ detection in the $t\anti t b\anti b$
\REF\hplus{J.F. Gunion, \plbj{322} (1994) 125.}
\REF\hplusroy{V. Barger, R.J.N. Phillips, D.P. Roy, \plbj{324} (1994) 236.}
final state.\refmark{\hplus,\hplusroy}
(The latter three channels require good efficiency ($\ebtag\sim0.3$)
and purity ($\emistag\lsim0.01$)
for vertex-tagging $b$-jets with $p_T\gsim 20\gev$ and $|\eta|\lsim 2$.)
In particular, it was demonstrated
in Ref.~[\dgvii] that $t\anti t \hl$ detection in the $t\anti t b\anti b$
channel is possible in the above-mentioned parameter space window,
and indeed guarantees that the LHC will find at least one of the MSSM
Higgs bosons. Similarly, $W\hl$ detection in the $W b\anti b$ mode
will be possible in this window, and perhaps to even higher $\mha$
values.\refmark\stangeetalwhsm\

In this paper, we focus on the region of large $\tanb$ (roughly $\gsim 5-10$).
For large $\tanb$ the $\ha$ and either the $\hh$ (for $\mha\gsim 110-150$,
depending upon $\mt$ and $\mstop$)
or the $\hl$ (for $\mha\lsim 110-150$) have highly enhanced couplings
to $b\anti b$ and $\tauptaum$. This has two important consequences:
i)
production rates via $gg\rta b\anti b \h$ can be greatly enhanced;
\Ref\willendicus{D. Dicus and S. Willenbrock, \prdj{39} (1989) 751.}\
and ii)
$\h$ decays will be dominated by $b\anti b$ ($BR\sim 0.9$) and
$\tauptaum$ ($BR\sim 0.1$).
Detailed simulations by
the LHC experimental groups\Ref\cmsatlas{D. Denegri (CMS)
and F. Pauss (ATLAS), private communications and seminar presentations.}
confirm the suggestion\refmark\kz\
that the $\hh$ and $\ha$ can be detected in the $\tauptaum$ decay
mode at high $\tanb$. Here, we analyze detection of $\h=\hl,\hh,\ha$
in the $b\anti b \h\rta b\anti b b\anti b$ final state. As in the
previously-considered final state modes involving $b$'s,
$b$-tagging will be critical.
\foot{Though not considered to date, we encourage the LHC experimental
groups to single $b$-tag in their $\tauptaum$ analyses; this should
greatly reduce their backgrounds and might increase the significance
of the signal.}
It is important to note that if $\tanb$ is large enough, then
the decays of the two $\h$'s with enhanced
$b\anti b,\tauptaum$ couplings will be dominated by the
$b\anti b,\tauptaum$ modes, even if sparticle-pair
channels are kinematically allowed. Thus,
at large $\tanb$ the $b\anti b\tauptaum$ and $b\anti b
b\anti b$ final states may provide the {\it only} access to two of the three
neutral MSSM Higgs bosons.

We require at least 3 jets with $p_T>20\gev$
and $|\eta|<2$, and that 3 or more jets
be tagged as $b$-jets (either by vertex separation or lepton-within-jet
techniques).  In the above fiducial range we assume
a tagging efficiency of $\sim 30\%$ (actually, we employ
the SDC detector vertex-tagging efficiency\Ref\sdcloi{Technical Design Report,
SDC Collaboration.}) and purity of 99\% (\ie\ a mis-tagging
probability of 1\%), provided a given $b$-jet
is separated by at least $\Delta R>0.5$ from neighboring jets.
To reveal the signal, we plot $M_{2b}$, where all pairs
of $b$'s and/or other jets, {\it both} of which are vertex-tagged
and separated from neighboring jets by $\Delta R>0.7$, are
included.

Aside from the combinatoric background from incorrect $2b$ combinations
in the signal reaction, there are the irreducible backgrounds
of $gg\rta b\anti b b\anti b$ and $gg\rta b\anti b Z\rta b\anti b b\anti b$.
The important reducible backgrounds are
$gg\rta b\anti b g$ with mis-tagging of the $g$ as a third $b$-jet,
and $t\anti t,t\anti t g$ backgrounds where a normal jet is mistagged
in addition to the two $b$'s deriving from the $t$ decays.
All processes have been computed with exact matrix elements. Further, we have
included the possibility that the $b$'s decay semi-leptonically
($b\rta cl\nu$).  This has been done at the quark level using the exact
decay matrix element.  For such decays,
the $c+l$ combination is presumed to form a
visible jet (subject to our usual fiducial cuts).

All jet, lepton momenta have been smeared (including the $c$ and
$l$ separately in semi-leptonic $b$ decays) using resolutions
$\Delta E/E=0.5/\sqrt E \oplus 0.03$, $0.2/\sqrt E \oplus 0.01$,
respectively. We have not incorporated
any reductions in our event rates due to detector efficiency.

With regard to QCD correction factors, we have included a $K$ factor
of 2 in both the $gg\rta b\anti b \h$ signal reaction and
in the $gg\rta b\anti b Z$ and
$gg\rta b\anti b b \anti b$ irreducible backgrounds.  Explicit
calculations of the actual $K$ factors are needed.  Experience with
other $gg$-induced reactions suggests that the assumed $K$ factors
are quite reasonable.  The $K$ factors employed roughly agree
with those estimated from the alternative analysis of the 3 $b$-tag
final state using the $g b\rta \h b$ plus
$g \anti b \rta \h \anti b$ signal processes
and $g b\rta Z b$ plus $g \anti b \rta Z \anti b$ and
$g b\rta b\anti b b$ plus $g \anti b \rta b \anti b \anti b$ irreducible
background subprocesses.\Ref\magnitude{See for example,
J.F. Gunion \etal, \npbj{294} (1987) 621;
F. Olness and W.-K. Tung, \npbj{308} (1988) 813;
D.A. Dicus and S. Willenbrock, \prdj{39} (1989) 751.}
The latter analysis was not employed here since it has the disadvantage
of not allowing a full tagging treatment for all the $b$ quarks in
the final state.
Regarding the reducible background processes, we note that the subprocess
$gg\rta b\anti b g$ is computed with a cutoff of
$30\gev$ on the $g$ transverse momentum.  This cutoff is chosen
so that the $gg\rta b\anti b g$ total cross section is of
similar magnitude to the $gg\rta b\anti b$ total cross section
computed at the born level, thus providing an effective $K$ factor of
order 2 for the $gg\rta b\anti b$ process including radiative corrections.
Clearly, this too is an approximate procedure.
Exactly the same procedure was employed
for the $gg\rta t\anti t $ and $gg\rta t\anti t g$ processes
in earlier work,\refmark{\dgvi,\hplus}\ and will again be used
for these processes here. Finally,  we note that
the gluon distribution functions we have employed are the $D0^\prime$
distributions of the MRS collaboration,
\Ref\mrs{A.D. Martin, R.G. Roberts, and W.J. Stirling,
\plbj{306} (1993) 145; Erratum, \ibid, {\bf B309} (1993) 492.}
and that $\alpha_s$ has been evaluated using the corresponding
$\Lambda_{QCD}$ value at a $Q$ specified by the
subprocess center-of-mass energy.

Another important QCD correction associated with the signal process
is the running of the $b$ quark mass. For $\mb(\mb)=4.5\gev$,
the $\mb(\mh)$ that appears in the
$b\anti b \h$ coupling is typically substantially smaller, falling
to as low as $\sim 3.1\gev$ at $\mh=450\gev$.  This running
mass is incorporated in computing both the $b\anti b\h$ cross section
and the $\h\rta b\anti b$ branching ratio.

 \TABLE\lhclum{}
 \topinsert
 \titlestyle{\twelvepoint
 Table \lhclum: We tabulate as a function of Higgs mass, $\mh$:
$\nsd=S/\sqrt B$ as computed for SM $\h b\anti b$ coupling and
$BR(\h\rta b\anti b)=1$; the mass interval employed, $\Delta M_{2b}$;
the {\it amplitude} enhancement factor, $E$,
required for a $5\sigma$ confidence
level signal in a $L=100\fbi$ LHC year;
the associated signal rate, $S(E)$; and the background rates
in this same mass bin --- $C(E)$ (for combinatoric background from the signal
reaction); $b\anti b b \anti b$; $b\anti b g$; $b\anti b Z$; $t\anti t g+
t\anti t$ at $\mt=140\gev$ (the $t\anti t$ backgrounds decrease with
increasing $\mt$). Event rates are obtained by summing
over the mass acceptance $\Delta M_{2b}$ specified for each $\mh$ value
and are in units of 1000 events.}
 \bigskip

 \thicksize=0pt
 \hrule \vskip .04in \hrule
 \begintable
 $\mh$ | $\nsd(E=1)$ | $\Delta M_{2b}$
| $E$ | $S(E)$ | $C(E)$ | $b\anti b b\anti b$ |
  $b\anti b g$ | $b\anti b Z$ | $t\anti t g+t\anti t$ \cr
  50 | 0.115 | 30 | 6.5  | 16.5 | 24.8 | 7680 | 3170 | 48.3 | 52.4        \nr
 100 | 0.078 | 30 | 8.1  | 13.4 | 21.3 | 3850 | 3170 | 57.6 | 68.8        \nr
 150 | 0.027 | 30 | 12.8 | 8.6  |  9.1 | 1550 | 1390 | 14.3 | 21.5        \nr
 200 | 0.021 | 30 | 14.7 | 5.6  |  4.3 |  657 |  584 |  5.9 |  9.1        \nr
 250 | 0.018 | 40 | 16.2 | 4.1  |  2.6 |  320 |  346 |  4.4 |  4.9        \nr
 350 | 0.010 | 60 | 23.4 | 2.6  |  2.3 |  145 |  129 |  1.8 |  1.3        \nr
 450 | 0.0063| 60 | 27.5 | 1.5  |  0.9 |   44 |   45 |  0.6 |  0.3   \endtable
 \hrule \vskip .04in \hrule
 \endinsert

\smallskip
\noindent{\bf 2. Results}
\smallskip

In Table~\lhclum\ we give results (at various $\mh$ values
\foot{Interpolation/extrapolation will be employed to obtain results
for other values of Higgs mass.})
for the number of standard deviations,
defined as $\nsd=S/\sqrt B$, that can be achieved at the LHC
with $\sqrt s=14\tev$ and integrated luminosity of $L=100\fbi$
assuming that the $b\anti b \h$ coupling is of SM strength and
that $BR(h\rta b\anti b)=1$.
\foot{The results are insensitive (within the statistics of
our Monte Carlos) to whether the $\h$ has scalar or pseudoscalar
fermion coupling.}
Here $S$ is the signal event rate (for an appropriately
chosen mass acceptance) obtained after subtracting the combinatoric
background using a smooth fit to neighboring bins.
$B$ includes the combinatoric background
subtracted from the signal, as well as all the backgrounds
explicitly outlined earlier.
Higgs signals and backgrounds were integrated
over the mass acceptances $\Delta M_{2b}$ tabulated in the third column.
Of course, the resulting nominal $\nsd$ values are small and it is only
after the signal enhancement rescalings appropriate to the MSSM at large
$\tanb$ that observable signals can be obtained.


To further quantify the nature of our results, we also give in Table~\lhclum\
the enhancement factor $E$ by which our standardized
signal must be multiplied {\it at the amplitude level} (\ie\ the cross
section is proportional to $E^2$) in order to obtain $\nsd=5$
for $L=100\fbi$.  Also given is the corresponding signal rate, $S(E)$,
the signal combinatoric background rate $C(E)$, as
well as the various background process rates.
{}From this table we see that
by far the biggest of the backgrounds are $gg\rta b\anti b b\anti b$
and $gg\rta b\anti b g$ production with mis-tagging of
the $g$. These are roughly comparable for our assumed
$\emistag=0.01$. Worsening the purity (\ie\ increasing $\emistag$)
would quickly increase the $b\anti b g$ background; a vertex detector
with excellent purity is very crucial to our signals.
Improvements in $\emistag$  to a level below 0.01 would only clean
up the signal somewhat, since the irreducible $b\anti b b\anti b$
background would not be affected. The combinatoric signal
background, though generally comparable to the signal,
is insignificant in comparison to the irreducible $gg\rta b\anti b
b\anti b$ background for $\tanb$ values such that
the signals first exceed $\nsd\sim 4-5$. (Of course, for {\it extremely}
large $\tanb$ values, the only background is the combinatoric
background from the signal.)
Relative to the $gg\rta b\anti b g$ and $gg\rta b\anti b b\anti b$ backgrounds,
the $t\anti t,t\anti t g$ backgrounds are neglible (for $\mt\gsim 140\gev$).
The $gg\rta b\anti b Z$ background can be neglected except
for $M_{2b}\sim \mz$ where it creates a $M_{2b}$ bump
somewhat larger than an $\nsd=5$ Higgs signal.
However, since the $Z$ bump normalization would be fairly accurately known
(using \eg\ the $b\anti b \lplm$ channel) it could be subtracted from the data.
Certainly a Higgs bump excess comparable to the $Z$ bump would be
quite apparent.  Thus, the actual $E$ needed for $\mh\sim\mz$
is probably somewhat above 7 but below 14.

For  poorer jet resolution, $\Delta E/E=0.8/\sqrt E\oplus0.05$,
the nominal $\nsd$ and corresponding $E$ values
become: $\nsd(E)=0.12(6.4)$, 0.081(7.8), 0.025(13.5),
0.019(16.4), 0.011(21.8), 0.0076(25.6), 0.0055(30.0) for the
seven Higgs masses, respectively.
Comparing to the $E$ values listed in Table~\lhclum,
we see that while good resolution
is helpful, at least at large $\mh$, it is not so vital
as high $b$-tagging efficiency and purity.

The Higgs bumps in the $2b$ mass spectrum that yield a nominal value
of $\nsd(E)=5$ typically have $S(E)/B\lsim 0.01$. To detect such bumps in
practice requires care, especially if $\mh$ is in the vicinity of the broad
peak in the background near $M_{2b}\sim 60 \gev$.
To convince ourselves that such nominal
$\nsd=5$ bumps can be found, we performed the following `blind' test.
We generated data in 5 GeV bins corresponding
to pure background, allowing bin by bin fluctuations
corresponding to the expected statistics. Data sets including
a nominal $\nsd=5$ Higgs bump were also generated.
For each data set we chose a large set of test Higgs masses, $\mh^0$.
For each $\mh^0$, we fit points of the given data set {\it outside}
the interval $[\mh^0-\half\Delta M_{2b},\mh^0+\half\Delta M_{2b}]$ to the form
$n(M_{2b})=B(M_{2b})\times
[c_0+c_1(M_{2b}-\mh^0)+c_2(M_{2b}-\mh^0)^2]$,\foot{In practice, all that is
needed is any form that provides a highly accurate and smooth fit.}
where $B(M_{2b})$ is our predicted background and $\Delta M_{2b}$
was varied with $\mh^0$ according to Table~\lhclum.
(To the extent that the normalization and shape deviations from a Monte Carlo
background prediction due to experimental effects can be quadratically
parametrized, the constants $c_{0,1,2}$
would automatically adjust to the normalization and shape of the actual
experimental data.)
For each test $\mh^0$, we then subtracted the fit from the data.
For a data set with a Higgs bump, an incorrect choice of test $\mh^0$
would yield a difference near the Higgs bump that was at first negative,
then positive, and again negative (as one would expect when trying
to fit a bump with a smooth function).  For a choice of $\mh^0$
near the true $\mh$, the difference result exhibited
a clear positive bump in the $\mh^0\pm \half\Delta M_{2b}$ interval.
The difference bump typically had a statistical significance of $4\sigma$.
We never failed to find
the Higgs signal, and for all the pure background samples generated
the largest false signal found was at the level of $2\sigma$.
Thus, assuming that there are no experimental sources of sharp $M_{2b}$
structure on a 30 GeV mass scale, we are convinced that low $S/B$
bumps with nominal $\nsd=5$ can be found.

We now turn to the MSSM Higgs bosons.
To convert the $\nsd(E=1)$ results of Table~\lhclum\ for use in the MSSM,
we multiply by $E^2=f(\h) BR(\h\rta b\anti b)$,
\foot{Significant $\mt$ dependence
enters indirectly through this factor, due to large Higgs sector
radiative corrections that depend on $\mt$.}
where $f(\h)=[\sina/\cosb]^2$, $[\cosa/\cosb]^2$, and $[\tan\beta]^2$ for
$\h=\hl$, $\hh$ and $\ha$, respectively, and
$BR(\h\rta b\anti b)$ is computed including all SM and SUSY decay modes.

\REF\erice{J.F. Gunion, `Detecting the SUSY Higgs Bosons',
Proceedings of the 23rd INFN Eloisatron Workshop on {\it Properties of
SUSY Particles}, eds. L. Cifarelli and V. Khoze, World Scientific
Publishing, p. 279.}

\FIG\surveybbbbi{
Discovery contours (at the $4\sigma$ level) in $\mha$--$\tanb$
parameter space for $\ebtag=0.3$ and $L=100\fbi$ at the LHC
for the reactions $t\anti t \h$ [h) $\h=\hh$; i) $\h=\hl$;
j) $\h=\ha$] and $b\anti b\h$ [k) $\h=\hh$; l) $\h=\hl$;
m) $\h=\ha$].
The contour corresponding to a given reaction is labelled by
the letter assigned to the reaction above. In each case, the letter
appears on the side of the contour for which detection of the
particular reaction {\it is} possible.
We have taken $\mt=175\gev$, $\mstop=1\tev$ and neglected squark mixing.
Scenario (A) refers to the notation established in Ref.~[\erice]; it
corresponds
to the case in which charginos and neutralinos are taken to be heavy.}

\midinsert
\vbox{\phantom{0}\vskip 8in
\phantom{0}
\vskip .5in
\hskip -97pt
\special{ insert user$1:[jfgucd.dai_bbbb]contour_bbbb_hbb_lhc_a_mt175.ps}
\vskip -1.95in }
{\rightskip=3pc
 \leftskip=3pc
 \Tenpoint\baselineskip=12pt
\noindent Figure~\surveybbbbi:
Discovery contours (at the $4\sigma$ level) in $\mha$--$\tanb$
parameter space for $\ebtag=0.3$ and $L=100\fbi$ at the LHC
for the reactions $t\anti t \h$ [h) $\h=\hh$; i) $\h=\hl$;
j) $\h=\ha$] and $b\anti b\h$ [k) $\h=\hh$; l) $\h=\hl$;
m) $\h=\ha$].
The contour corresponding to a given reaction is labelled by
the letter assigned to the reaction above. In each case, the letter
appears on the side of the contour for which detection of the
particular reaction {\it is} possible.
We have taken $\mt=175\gev$, $\mstop=1\tev$ and neglected squark mixing.
Scenario (A) refers to the notation established in Ref.~[\erice]; it
corresponds
to the case in which charginos and neutralinos are taken to be heavy.}

\endinsert

We present our results using the conventional $\mha$--$\tanb$ parameter space.
We consider first the standard reference scenario in which the Higgs radiative
corrections\Ref\perhaber{For a review, see H.E. Haber,
{\it Perspectives on Higgs Physics}, ed. G. Kane,
World Scientific Publishing (1992), p. 79.}\
are specified by $\mstop=1\tev$, with neglect of squark mixing, and
all SUSY particles are assumed too heavy to appear in Higgs decays.
The $\nsd=4$ discovery contours are shown in Fig.~\surveybbbbi\
for $\mt=175\gev$. Also displayed in these figures are the
$\nsd=4$ contours for the $t\anti t \h$ discovery modes
considered in Ref.~[\dgvii]. Figure~\surveybbbbi\ shows that
discovery of the $\ha$ in the $b\anti b b\anti b$
mode will be possible in the triangular
wedge of parameter space lying roughly above a line beginning at $\tanb\sim 5$
at low $\mha$ rising to $\tanb\sim 20$ at $\mha\sim 350\gev$.
For $\mha\lsim 125\gev$, $\hl$ discovery will be possible above this line
(it is the $\hl$, and not the $\hh$, which has enhanced $b\anti b$ couplings
at large $\tanb$ for $\mha\lsim 125\gev$), while for $\mha\gsim 125\gev$
the $\hh$ will be visible for $\tanb$ values above this line. In other words,
we can detect either the $\ha$ and $\hl$ (for $\mha\lsim 125\gev$)
or the $\ha$ and $\hh$ (for $\mha\gsim 125\gev$), provided
$\tanb$ exceeds the necessary value. Thus, this discovery
mode could be very crucial for MSSM GUT models which require large $\tanb$,
such as models in which the top, bottom and tau Yukawa couplings are
all required to have the same GUT value. Fig.~\surveybbbbi\ also
shows how nicely the $b\anti b b\anti b$ mode complements the
$t\anti t b\anti b$ mode at $\mt=175\gev$.
(The viable region for the $W b\anti b$ mode is similar to, but somewhat more
extensive than shown for the $t\anti t b\anti b$ mode.\refmark\stangeetalwhsm)
Combining all $b$-tagging modes, much of parameter space
is covered in one way or another.  And certainly, the parameter
space window/hole mentioned in the introduction
(see Ref.~[\dgvii], for example, for a figure showing the hole)
is now covered by at least one and usually two or three different
$b$-tagging related modes.

As $\mt$ moves lower (higher) the $\hl$--$\hh$ boundary for the $b\anti b b
\anti b$ modes shifts to lower (higher) $\mha$ values (\eg\ $\mha\sim 110\gev$
for $\mt=150\gev$ and $\mha \sim 140\gev$ for $\mt=200\gev$), but
the overall region covered by the $b\anti b b\anti b$ modes remains
more or less unchanged. Also, the $Wb\anti b$ mode coverage remains more or
less
unchanged, whereas the $t\anti t b\anti b$ mode coverage expands
(contracts). (In fact, for $\mt=200\gev$ the $t\anti t b\anti b$
modes do not reach the $\nsd=4$ criterion for very much of parameter
space under the luminosity \etc\ assumptions made in our analysis.)
Meanwhile, the `hole' that needs to be covered increases (decreases) in size,
disappearing altogether by $\mt=200\gev$, at which value
at least one of the MSSM Higgs bosons can be found using the $\gam\gam$
(or $l\gam\gam$) and $4l$ Higgs
search modes.\Ref\gunorr{See J.F. Gunion and L. Orr, Ref.~[\everybody].}\
In combination, the $b$-tagging modes allow discovery of at least
one MSSM Higgs boson whenever there is a hole in $\mha$--$\tanb$
parameter space where {\it both} the $\hh$ and $\hl$ cannot be found in either
the $\gam\gam$ or $4l$ final states.

\FIG\surveybbbbiii{
Discovery contours at the $4\sigma$ level for $\ebtag=0.4$ and $L=200\fbi$.
We have taken $\mt=175\gev$, $\mstop=1\tev$ and neglected squark mixing.}
\midinsert
\vbox{\phantom{0}\vskip 8in
\phantom{0}
\vskip .5in
\hskip -97pt
\special{ insert
user$1:[jfgucd.dai_bbbb]contour_bbbb_hbb_lhc_a_2sigma_mt175.ps}
\vskip -1.95in }
{\rightskip=3pc
 \leftskip=3pc
 \Tenpoint\baselineskip=12pt
\noindent Figure~\surveybbbbiii:
Discovery contours at the $4\sigma$ level for $\ebtag=0.4$ and $L=200\fbi$.
We have taken $\mt=175\gev$, $\mstop=1\tev$ and neglected squark mixing.
Notation as in Fig.~\surveybbbbi}
\endinsert

It is important to ask how sensitive these discovery contours
are to our input luminosity and $b$-tagging efficiency and purity.
Increases in $\ebtag$ above 0.3 are most effective, provided purity at
$\emistag=0.01$ can be maintained at the same time.  It seems
possible that $\ebtag=0.4$ could be achieved after
combining vertex and lepton-decay tagging.  Meanwhile, after several
years of running an integrated luminosity of $L=200\fbi$ might
be feasible.  In combination, this would more than quadruple the number
of events in both the signal and irreducible backgrounds (somewhat less
for reducible backgrounds keeping fixed $\emistag=0.01$). The results obtained
for $\ebtag=0.4$ and $L=200\fbi$ if we require $\nsd=4$ are depicted
in Fig.~\surveybbbbiii. The $b$-tagging modes provide dramatic
coverage of all of parameter space --- the $b\anti b b\anti b$
modes cover a wedge from $\tanb\sim 4$ at small $\mha$
to $\tanb\sim 15$ at $\mha\sim 400\gev$, while the $t\anti t b\anti b$\
mode for the $\hl$ ($\hh$) covers $\mha\gsim 120\gev$ for all $\tanb$
($60\lsim\mha\lsim 120\gev$ for $\tanb\gsim 2.5$).
Indeed, one of the two $\hl$
modes --- $t\anti t b\anti b$ [i)] and $b \anti b b\anti b$ [l)] ---
is viable for almost all of parameter space.
The $\ebtag=0.4$, $L=200\fbi$ ($\nsd=4$) contours for $\mt=150$ or $200\gev$
are
very similar, with the $\hl$ discoverable in either the
i) or l) mode in almost all of $\mha$--$\tanb$ parameter space.

How do these results change if $\mstop$ is reduced and/or sparticle
pair decay modes are allowed for the heavier Higgs bosons?
If $\mstop$ is reduced, the radiative corrections to the Higgs masses
become smaller.  The primary effect for the $b\anti b b\anti b$ mode
is to shift the boundary between the $\hl$ and $\hh$ regions
and the neighboring `flex' point in the $\ha$ boundary to
somewhat lower $\mha$ (\eg\ for $\mt=200\gev$,
from $\mha\sim 140\gev$ to $\mha\sim 100\gev$ if
$\mstop$ is reduced from $1\tev$ to $300\gev$).
Since the $b\anti b b\anti b$ modes are only viable when $\tanb$ is large
and the couplings to $b\anti b,\tauptaum$ greatly enhanced,  SUSY decay
modes cannot generally compete with the $b\anti b$ and $\tauptaum$
decay modes and, therefore, do not influence branching ratios
unless the ino and slepton masses are {\it very} light.
Even in such scenarios, the $b\anti b$ branching ratio is only reduced
to $0.5-0.6$ (vs. the nominal 0.9) for $\tanb$ values that lie just within
the discovery regions. Meanwhile,
the $t\anti t b\anti b$ modes are viable when
the Higgs bosons are light and have roughly SM couplings ---
this means $t\anti t \hl$ is viable at large $\mha$ where the $\hl$
becomes SM-like and is never particularly heavy,
and $t\anti t \hh$ is only viable
when $\mha$ is relatively small so that the $\hh$ is relatively
light and has SM-like couplings. Since the Higgs
in question is always light ($\lsim 130\gev$) SUSY decay modes
are unlikely to be important (the possible exception being
invisible $\cnone\cnone$ and $\snu\snu$ decays
\REF\tthinvisible{J.F. Gunion, \prlj{72} (1994) 199.}
\REF\whinvisible{S.G. Frederiksen, N.P. Johnson,
G.L. Kane, and J.H. Reid, preprint SSCL-577-mc (1992);
D. Choudhury and  D.P. Roy, \plbj{322} (1994) 368.}\
which can be detected in the $t\anti t+\etmiss$\refmark\tthinvisible\
and $W+\etmiss$\refmark\whinvisible\ modes). The primary effect upon
the $t\anti t b\anti b$ mode of reducing $\mstop$ is to decrease the
maximum $\hl$ mass, thereby further emphasizing
the importance of its $b\anti b$
decay modes at large $\mha$ --- for example, the region labelled by i) in
Fig.~\surveybbbbi\ expands greatly for $\mstop\sim 300\gev$.

\smallskip
\noindent{\bf 3. Conclusion}
\smallskip

We have demonstrated that $b$-tagging can be used to isolate
$b\anti b \h$ ($\h=\hl,\hh,\ha$) events, in which $\h\rta b\anti b$,
over much of the large-$\tanb$ portion of the Minimal
Supersymmetric Model $\mha$--$\tanb$ parameter space.
The $t\anti t \h,W\h$ ($\h\rta b\anti b$) modes are highly complementary
in that they allow detection of one of the neutral MSSM Higgs bosons
for a large part of the rest of the parameter space, while also
overlapping (in part) the $b\anti b b\anti b$ mode coverage.
Of course, an accumulated luminosity of $L=100\fbi$
is needed and $b$-tagging must be possible with good efficiency
($\ebtag\gsim 0.3$) and good purity ($\emistag\lsim 0.01$). Parameter
space coverage expands dramatically with relatively small
improvements in $L$ and, especially, $\ebtag$. For $L=200\fbi$
and $\ebtag\gsim 0.4$, discovery of all of the
neutral MSSM Higgs bosons becomes possible for $\tanb$ above the
$\mha$-dependent minimum value at which the $b\anti b b\anti b$ modes
become viable. [For large $\mha$ (\eg\  $\mha\gsim 125\gev$
if $\mt\sim 175\gev$)
the $\hl$ would be found in the $t\anti t b\anti b$ and $Wb\anti b$ final
states, and the $\hh$ and $\ha$ in the $b\anti b b\anti b$ mode.]
For very large $\tanb$ ($\gsim 30$), charged Higgs detection
in the $t\anti b b\anti t$ mode becomes possible and all the MSSM
Higgs bosons could be found at the LHC in $b$-tagged modes.
These results are more or less independent of
other MSSM parameters such as $\mstop$ and the ino and slepton mass scales.
The very great promise of the $b$-tagging modes,
and the fact that, at a minimum, they guarantee
that the LHC alone will find at least one
of the MSSM Higgs bosons (for any value of $\mt\gsim 140\gev$)
makes it virtually mandatory that the detector collaborations at the
LHC find a way to perform $b$-tagging with the required efficiency
and purity in the multi-event-per-crossing environment that
will prevail for high instantaneous luminosity at the LHC.

\smallskip\noindent{\bf 4. Acknowledgements}
\smallskip
This work has been supported in part by Department of Energy
grants \#DE-FG03-91ER40674, \#DE-FG03-90ER40546, and \#DE-FG05-92ER40722,
and by grants \#RGFY-93-330 and \#RCFY-93-229 from the
Texas National Research Laboratory.
JFG would like to thank D. Denegri and M. Della Negra
for illuminating conversations regarding the difficulties of $b$ tagging
at high luminosity.

\smallskip
\refout
\end